\documentclass[11pt]{article}%
\usepackage{cite}
\usepackage{amssymb}
\usepackage{amsmath}
\usepackage{bm}%
\usepackage{amsfonts}%
\usepackage{graphicx}
\providecommand{\U}[1]{\protect\rule{.1in}{.1in}}
\begin{document}

\title{Quantum Trajectories: Dirac, Moyal and Bohm.}
\author{B. J. Hiley$^{1}$, M. A. de Gosson$^{2}$ and G. Dennis$^{1}$.}
\date{$^{1}$ Physics Department, University College, London, Gower Street, London
WC1E 6BT.\\
$^{2}$ Universit\"{a}t Wien, NuHAG, Fakult\"{a}t f\"{u}r Mathematik, \\
A-1090 Wien.}

\maketitle

\begin{abstract}
We recall Dirac's early proposals to develop a description of quantum
phenomena in terms of a non-commutative algebra in which he suggested a way to
construct what he called `quantum trajectories'. Generalising these ideas, we
show how they are related to weak values and explore their use in the
experimental construction of quantum trajectories. We discuss covering spaces
which play an essential role in accounting for the `wave' properties of
quantum particles. We briefly point out how new mathematical techniques take
us beyond Hilbert space and into a deeper structure which connects with the
algebras originally introduced by Born, Heisenberg and Jordan. This enables us
to bring out the geometric aspects of quantum phenomena.

\end{abstract}

\section{Introduction}

In a classic paper, Dirac\cite{pd45} has drawn attention to the similarity of
the \emph{form}
of the classical dynamical equations expressed in terms of commuting functions
and the \emph{form} of the corresponding non-commutative operator equations
appearing in the quantum domain. The latter, essentially Heisenberg mechanics,
can be represented by matrices and therefore form part of a non-commutative
algebraic structure. This is in contrast to the Schr\"{o}dinger approach which
is represented in a formal Hilbert space structure, and leads to more familiar
mathematics based on differential operators acting on continuous wave
functions, the non-commutativity being taken care of in the form of the
differential operators. These techniques, being more familiar to physicists,
quickly generated results and placed the Schr\"{o}dinger picture in prime
position. This has led to the conclusion that the quantum `particle' appears
more wave-like than the particles of classical dynamics.

In spite of this, Dirac felt that the replacement of commuting functions by
non-commuting variables pointed to a deeper connection between the algebraic
approach and classical mechanics and suggested that this relationship should
be examined more closely. In making this proposal he realised that techniques
necessary for handling non-commuting mathematics were not readily available.
Nevertheless Dirac made some tentative suggestions on how to construct quantum
expectation values when general non-commuting variables were involved. With
these techniques at hand, he attempted to generalise the notion of a contact
transformation to the quantum situation. Dirac thereby provided a method of
constructing what he called the ``trajectories of a quantum particle" based on
a non-commutative structure and without using wave functions explicitly.

However these attempts were soon superseded by a third approach, the path
integral method, which was proposed by Feynman\cite{rf49}
after he had read Dirac's paper. With the success of this approach, the notion
of an actual quantum trajectory was dropped, particularly as Mott\cite{nm29}
had shown how the wave equation could be used to explain the trajectories seen
in particle detectors like cloud chambers. This, together with the uncertainty
principle, discouraged any further consideration
of particle trajectories in the quantum domain. Moreover since no operational
meaning could be given to such a notion, further discussion ceased. Thus
Dirac's idea of constructing quantum trajectories was abandoned and even forgotten.

In the meantime, the more general debate concerning the completeness of the
quantum formalism, initiated by Einstein, Podolsky and Rosen\cite{aebpnr35},
continued unabated, focusing on the possibility of adding `hidden variables'
and thereby allowing for the possibility of trajectories. This was in spite of
von Neumann's\cite{jvn55}
claim to have proved that such variables could not be used to explain the
statistical properties of quantum processes without contradicting experimental
results. However in 1952 a paper by Bohm\cite{db52}
appeared claiming that by simply splitting the Schr\"{o}dinger equation into
its real and imaginary parts, a more detailed account of quantum phenomena
could, in fact, be given based on particle trajectories.

Unfortunately the phrase `hidden variables' was used in the title of the paper
whereas Bohm actually introduced no additional parameters at all into the
formalism. He had merely interpreted the existing formalism in a novel way. In
fact he had simply shown that the real part of the Schr\"{o}dinger equation,
under polar decomposition of the wave function, was of a form that looked
remarkably like the classical Hamilton-Jacobi equation provided certain
relations valid in the classical domain could be extended into the quantum
domain. This equation, which we call the quantum Hamilton-Jacobi equation,
enabled the straightforward calculation of what appeared to be `trajectories'
as was demonstrated by Philippidis \emph{et al}. \cite{cdcpbh79} for an
ensemble of
particles constrained by certain experimental conditions such as defined in,
for example, the two-slit experiment. Explanations of other quantum phenomena,
again in terms of these `trajectories' followed, giving rise to an alternative
understanding of these phenomena in a way that was thought to be impossible.
(See Bohm and Hiley\cite{dbbh93} and Holland\cite{ph94}.)

Thus contrary to expectation, these calculations demonstrated that it was
possible to account for the interference phenomena in terms of collections of
individual particle trajectories, although a deeper analysis raised the
question of exactly what meaning could be given to the notion of a quantum
particle following a trajectory. Unfortunately there seemed no way of
experimentally determining these trajectories and so they remained a curiosity
without experimental meaning. However some did embrace these ideas and
developed a topic called `Bohmian mechanics'\cite{ddst09}, using concepts that
Bohm himself did not enthusiastically embrace, the latter arguing that
something deeper was involved\cite{db68}. In this paper we will continue to
call these flow lines `trajectories'.

An examination of the two-slit experiment shows that the trajectories are not
straight lines after they pass through the slits even though no classical
potentials exist. The cause of these deviations could immediately be traced to
the presence of the extra term appearing in the quantum Hamilton-Jacobi
equation. At first, it was thought this extra term was merely an additional
new classical potential since without it the particles would move in straight
lines and no `fringes' would appear.

However a closer examination showed it to be very different from any known
classical potential. It had no external point source; it was non-local,
accounting for the effects of quantum entanglement and it reflected the
properties of the immediate experimental arrangement, adding support to Bohr's
notion of wholeness which he emphasised by demanding that the experimental
conditions be included in the description. In many ways it seemed to be a new
form of \emph{inner energy} possessed by the particle, organising the flow
lines in a novel way and suggesting a `formative' cause rather than the traditional efficient
cause~\cite{dbbh93} (also see \cite{degohi14,degohi15}).

Unfortunately the inclusion of the phrase `hidden variables' in Bohm's paper,
led to the belief that this was an attempt to return to a classical view of
the world based on the old notion of mechanics, in contrast to the dominant
view which was that such a return was impossible and a much more radical
outlook was required. Bohm agreed and simply considered his proposal as a
preliminary one providing a way to open up other, deeper possibilities.

However in the rather toxic atmosphere of the time, it was not realised that
Bohm had added nothing new to the mathematical structure and was merely
exploring the full implications of the quantum formalism in a different way.
It should not be forgotten that the striking result of this approach was to
bring out the notion of non-locality in entangled systems. Indeed it was
Bohm's work that prompted Bell\cite{jb64} to explore the wider
consequences of this non-locality. Thus, far from returning to a classical
picture, Bohm's work showed that the formalism contained many features that
were clearly not classical and the whole approach was actually pointing to a
radically new outlook.

Superficially, however, the Bohm approach did look naive as it provided no
connection with the Heisenberg approach, not only in the sense that it seemed
to violate the uncertainty principle, but it also seemed to avoid completely
the non-commutative properties of the Heisenberg algebra. Rather than trying
to understand how this approach produced results that were consistent with
those deduced from the non-commuting operators, the discussion degenerated
into a quasi-ideological battle between the two opposing views that emerged
from \emph{exactly the same mathematical structure}.

However a recent paper\cite{bh15} pointed out that a
non-commutative Heisenberg algebra had been further developed by von Neumann
who showed how quantum phenomena emerged from a non-commutative phase space.
This algebra was rediscovered by Moyal\cite{jm49} who demonstrated that this
approach could be understood as a generalisation of classical statistics to a
new kind of statistical theory that was demanded by non-commutativity. Carried
further, this non-commutativity seemed to require two time-dependent evolution
equations~\cite{bh15}. In the Moyal algebra, for example, one of these is
based on the Moyal bracket and the other on the Baker bracket\cite{gb58}. In
the classical limit,
the first of these equations becomes the Liouville equation. While the second,
based on the Baker bracket, reduces to the classical Hamilton-Jacobi equation.
These two equations have an operator analogue based on the commutator and the
anti-commutator, or Jordan product, which will be discussed in detail in
section \ref{sec:timedev}. When these equations are projected into the
$x$-representation they become the quantum Liouville equation and the quantum
Hamilton-Jacobi equation respectively. This immediately shows that the
equations defining the Bohm approach are projections from a non-commutative
space onto a shadow commutative phase space. (For a detailed discussion see
Hiley\cite{bh15}.)

There is one further connection between the Moyal approach and the Bohm
approach that is important to point out at this stage. The so-called guidance
condition, $P_{B}=\nabla S$, also known as the Bohm momentum, which enables
the direct calculation of the quantum trajectories, turns out to be the
conditional momentum given by the Moyal joint distribution function $\hat
{f}(\hat{X},\hat{P})$. Here $(\hat{X},\hat{P})$ are the operator equivalents
of the coordinates of a cell in phase space, the so called `quantum
blob'\cite{mdg13} although a deeper mathematical explanation
exists~\cite{Varilly}, which we briefly introduce in section \ref{sec:3.2}.
Furthermore as we have
already pointed out, one of the conditional time-development
equations is identical to the quantum Hamilton-Jacobi equation, becoming the
classical Hamilton-Jacobi equation in the appropriate limit. Thus the von
Neumann-Moyal approach, based on a non-commutative algebra and the Bohm model
are much more closely related than generally realised. In fact it could be
argued that the Bohm approach forms an integral part of Heisenberg's matrix
mechanics providing an intuitive account of the approach.

This brings us full circle to a classic paper by Dirac \cite{pd45} which calls
for a further investigation of the non-commutative Heisenberg approach. As we
have already indicated, Dirac constructed a general distribution function for
$n$ non-commuting variables, which for the special case of two variables
reduces to the Moyal distribution referred to above. Unfortunately Dirac
incorrectly thought that Moyal's theory only dealt with operators of the form
$e^{i(a\hat X+b\hat P)}$ whereas, in fact, this term was used to define a
distribution in phase space from which expectation values of \emph{any
function} of $(\hat X, \hat P)$ can be calculated. This distribution is
actually the Wigner function.

As has already been pointed out by one of us\cite{mdgmdg12}, the
cross-Wigner function can be identified with the weak value of the momentum
operator. In fact Dirac himself had implicitly introduced a weak value
although he did not give it that name and saw his work as an opportunity to
``discuss trajectories for the motion of a particle in quantum mechanics and
thus make quantum mechanics more closely resemble classical mechanics"--his
words, not ours\cite{pd45}.

\section{Dirac's Quantum Trajectories}

Regarding $\langle x_{t_{f}}| x_{t_{0}}\rangle$ as the probability amplitude
of a particle travelling from position $x_{t_{0}}$ to position $x_{t_{f}}$ and
travelling through a set of intermediate points, we can write
\begin{align*}
\langle x_{t_{f}}| x_{t_{0}}\rangle=\int\dots\int\langle x_{t_{f}}%
|x_{n}\rangle dx_{n}\langle x_{n}|x_{n-1}\rangle dx_{n-1}\dots dx_{2} \langle
x_{2}|x_{1}\rangle dx_{1}\langle x_{1}|x_{t_{0}}\rangle.
\end{align*}
where $\langle x_{i+1}|x_{i}\rangle$ is the propagator of the particle being
at $x_{i}$ at $t_{i}$ and arriving at $x_{i+1}$ at time $t_{i+1}$. Today we
would write this as
\begin{align*}
\langle x_{i+1}|x_{i}\rangle=\langle x_{i+1}|\exp[-i\hat H(t_{i+1}%
-t_{i})]|x_{i}\rangle
\end{align*}
but we will continue with the abridged notation for simplicity.

Thus a path is built up from a series of transitions between pairs of
neighbouring points, $(x_{i},x_{i+1})$ and the expectation value of an
operator during each transition is given by
\begin{align}
\langle x_{i+1}|\hat F(\hat X_{i}, \hat X_{i+1})|x_{i}\rangle=f(x_{i+1}%
,x_{i})\langle x_{i+1}|x_{i}\rangle\label{eq:A}%
\end{align}
where $f(x_{i+1}, x_{i})$ is the expectation value of the operator during the
transition $x_{i}\rightarrow x_{i+1}$. Furthermore we will assume the time
$\epsilon=(t_{i+1}-t_{i})$ to be small so that the trajectory can be divided
into infinitesimal segments. Clearly we can now regard the element $\langle
x_{i+1}|x_{i}\rangle_{\epsilon}$ as a propagator, which is written in the
form
\begin{align}
\langle x_{i+1}|x_{i}\rangle_{\epsilon}=\exp(iS_{\epsilon}(x_{i}%
,x_{i+1})/\hbar).\label{eq:DBprop}%
\end{align}

We will not, at this stage, identify the propagators with the Feynman
propagators although clearly they are related. We will regard $S_{\epsilon
}(x_{i},x_{i+1})$ as a function generating the motion. Then, taking the
momentum as an example, we find
\begin{align*}
\langle x_{i+1}|\hat P_{i+1}|x_{i}\rangle_{\epsilon}=-i\hbar\frac{\partial
}{\partial x_{i+1}}\langle x_{i+1}|x_{i}\rangle_{\epsilon}=\frac{\partial
S_{\epsilon}(x_{i}, x_{i+1})}{\partial x_{i+1}}\langle x_{i+1}|x_{i}%
\rangle_{\epsilon}\\
= \left\langle x_{i+1}\left|\frac{\partial S_{\epsilon}(x_{i}, x_{i+1})}{\partial
x_{i+1}}\right|x_{i}\right\rangle _{\epsilon}.%
\end{align*}
Thus from equation (\ref{eq:A}) we find
\begin{align}
p_{i+1}=\frac{\partial S_{\epsilon}(x_{i},x_{i+1})}{\partial x_{i+1}%
}.\label{eq:p1}%
\end{align}
Similarly we can consider
\begin{align*}
\langle x_{i+1}|\hat P_{i}|x_{i}\rangle_{\epsilon}=i\hbar\frac{\partial
}{\partial x_{i}}\langle x_{i+1}|x_{i}\rangle_{\epsilon}=-\frac{\partial
S_{\epsilon}(x_{i}, x_{i+1})}{\partial x_{i}}\langle x_{i+1}|x_{i}%
\rangle_{\epsilon}\\
=- \left\langle x_{i+1}\left|\frac{\partial S_{\epsilon}(x_{i}, x_{i+1})}{\partial
x_{i}}\right|x_{i}\right\rangle _{\epsilon}%
\end{align*}
so that
\begin{align}
p_{i}=-\frac{\partial S_{\epsilon}(x_{i},x_{i+1})}{\partial x_{i}%
}.\label{eq:p2}%
\end{align}
Dirac suggested that $p_{i}$ could be regarded as the momentum at the initial
point $(x_{i}, t_{i})$ of the interval while $p_{i+1}$ is the momentum at the
final point $(x_{i+1},t_{i+1})$, but clearly these are not eigenvalues of the
momentum operators, so what are they?

\subsection{The Classical Hamilton-Jacobi Theory}

\label{sec:chj}

Let us proceed cautiously, first by recalling that the formulae (\ref{eq:p1})
and (\ref{eq:p2}) are reminiscent of classical Hamilton-Jacobi theory. In this
theory the function, $S(x,x_{0};t,t_{0})$ generates a flow, or more
technically, a symplectomorphism, $f_{t,t_{0}}$, such that
\begin{equation}
(x,p)\longmapsto f_{t,t_{0}}(x_{0},p_{0}).\label{eq:flow}%
\end{equation}
These symplectomorphisms are elements of $\operatorname*{Ham}(2n)$, the group
of Hamiltonian symplectomorphisms \cite{RMP}. The flow is just another way to
write Hamilton's equations of motion
\begin{equation}
\frac{d}{dt}f_{t}(x,p)=X_{H}(x,p)
\end{equation}
where $X_{H}$ is the Hamiltonian vector field
\begin{equation}
X_{H}=\left(  \frac{\partial H}{\partial p},-\frac{\partial H}{\partial
x}\right)  .
\end{equation}
The time dependent flow is then defined as
\begin{equation}
f_{t,t_{0}}(x_{0},p_{0},t_{0})=f_{t-t_{0}}(x_{0},p_{0},t_{0})
\end{equation}
so that equation (\ref{eq:flow}) holds and defines functions $t\rightarrow
x(t)$ and $t\rightarrow p(t)$ satisfying
\begin{equation}
\dot{x}(t)=\frac{\partial H(x(t),p(t),t)}{\partial p};\quad\quad\dot
{p}(t)=-\frac{\partial H(x(t),p(t),t)}{\partial x}.
\end{equation}
The corresponding Hamilton-Jacobi equation is defined as
\begin{equation}
\frac{\partial S(x,x_{0})}{\partial t}+H\left(  x,\frac{\partial S(x,x_{0}%
)}{\partial x},t\right)  =0.
\end{equation}
Then for a free symplectomorphism $(x,p)=f_{t,t_{0}}(x_{0},p_{0})$, the
following relations must be satisfied
\begin{equation}
p=\frac{\partial S(x,x_{0};t,t_{0})}{\partial x};\quad\quad p_{0}%
=-\frac{\partial S(x,x_{0};t,t_{0})}{\partial x_{0}}.\label{eq:DiracPs}%
\end{equation}
A remarkable similarity with quantum equations (\ref{eq:p1}) and
(\ref{eq:p2})? Yes, but notice that the generating function for $f_{t,t_{0}}$
is $S(x,x_{0};t,t_{0})$ whereas the generating function for the quantum case
uses the \emph{exponential} of $S(x,x_{0};t,t_{0})$, namely, equation
(\ref{eq:DBprop}). This generates a different flow $F_{t,t_{0}}$, not in the
group $\operatorname*{Ham}(2n)$ but in its \emph{covering group}. In the
linear case (\textit{i.e.} when the Hamiltonian is quadratic), $f_{t,t_{0}}$
is an element of the symplectic group $\operatorname*{Sp}(2n)$, and
$F_{t,t_{0}}$ is an element of the metaplectic group $\operatorname*{Mp}(2n)$,
the double cover of the symplectic group. What one can show is that there is a
1--1 correspondence between the continuous curves $t\rightarrow f_{t,t_{0}}$
in $\operatorname*{Sp}(2n) $ and the continuous curves $t\rightarrow
F_{t,t_{0}}$ in $\operatorname*{Mp}(2n)$~\cite{mdgbh11}.

The reference to a covering group is not totally unknown in physics. The
notion of spin arises from a double cover, not of the symplectic group, but of
the orthogonal group. In the case of spin, the spin group is just the double
cover of the orthogonal group~\cite{bhbc12}. Similarly the
metaplectic group provides a double cover for the symplectic group. Properties
of both covering groups produce quantum effects that have been experimentally
demonstrated~\cite{akgo76, cc59, pced06}. So clearly the notion of a
covering space plays a key role in quantum mechanics.

The relation between the symplectic group and its double cover is provided by
the projection
\begin{equation}
\Pi:\operatorname*{Mp}(2n)\longrightarrow\operatorname*{Sp}(2n).
\end{equation}
Thus if $f_{t,t_{0}}$ is the flow determined by the generating function
$S(x,x_{0};t,t_{0})$ then, in the linear case
\begin{equation}
\psi(x,t)=F_{t,t_{0}}\psi(x_{0},t_{0})=A\int e^{iS_{t,t_{0}}(x,x_{0};t,t_{0}%
)}\psi(x_{0},t_{0})d^{3}x_{0}%
\end{equation}
where $A$ is a convenient normalisation factor (this formula remains true for
short times in the general case). One can show that $\psi(x,t)$ is a solution
of the Schr\"{o}dinger equation. For a complete account we need to extend the
covering group to $\operatorname*{Ham}(2n)$ which is the non-linear
generalisation of $\operatorname*{Sp}(2n)$. For a more detailed discussion see
de Gosson and Hiley~\cite{mdgbh11}.

\subsection{The Quantum Hamilton-Jacobi Equation}

Having noticed the similarity between the Dirac equations (\ref{eq:p1}) and
(\ref{eq:p2}) and the corresponding classical equations (\ref{eq:DiracPs}),
let us now try to exploit this similarity in a different way. We start
from equation (\ref{eq:p2}), which we write in a slightly simpler notation
as
\begin{align}
p_{0}=-\frac{\partial S_{\epsilon}}{\partial{x_{0}}}(x,x_{0};t,t_{0}%
).\label{eq:p0}%
\end{align}
We will now regard $(x, x_{0})$ as two independent variables. It follows from
the implicit function theorem that equation (\ref{eq:p0}) determines a
function $x=x^{\psi}(t)$ provided
\begin{align*}
\frac{\partial^{2} S_{\epsilon}}{\partial x\partial x_{0}}\ne0.
\end{align*}
We can then write
\begin{align}
p_{0}=-\frac{\partial S_{\epsilon}}{\partial{x_{0}}}(x^{\psi}(t),x_{0}, t,
t_{0}),\label{eq:PB0}%
\end{align}
where $x_{0}$ and $t_{0}$ are to be viewed as independent parameters. Then let
us define
\begin{align}
p^{\psi}(t)=\frac{\partial S_{\epsilon}}{\partial x}(x^{\psi}(t),x_{0}, t,
t_{0}).\label{eq:GC}%
\end{align}
The functions $x^{\psi}(t)$ and $p^{\psi}(t)$ can then be shown to be
solutions of the following Hamilton equations
\begin{align}
\dot x^{\psi}(t)=\frac{\partial H^{\psi}}{\partial p}(x^{\psi}(t),p^{\psi
}(t),t);\quad\quad\dot p^{\psi}(t)=-\frac{\partial H^{\psi}}{\partial
x}(x^{\psi}(t),p^{\psi}(t),t)\label{eq:HES}%
\end{align}
with the initial conditions $x^{\psi}(t_{0})=x_{0},\;p^{\psi}(t_{0})=p_{0}$;
here we have written our Hamiltonian as $H^{\psi}$ because it clearly cannot
be the classical Hamiltonian as that would not have produced any quantum
behaviour so what form will $H^{\psi}$ take?

The corresponding Hamilton-Jacobi equation now becomes
\begin{align}
\frac{\partial S}{\partial t}+H^{\psi}\left( x,\frac{\partial S}{\partial
x},t\right) =0.\label{eq:QHJ}%
\end{align}
To show that this equation is equivalent to the pair of equations
(\ref{eq:HES}), first differentiate (\ref{eq:QHJ}) with respect to $x_{0}$ and
find
\begin{align}
\frac{\partial^{2} S}{\partial x_{0}\partial t}+\frac{\partial H^{\psi}%
}{\partial x_{0}}= \frac{\partial^{2} S}{\partial x_{0}\partial t}%
+\frac{\partial H^{\psi}}{\partial p}\frac{\partial^{2}S}{\partial
x_{0}\partial x}=0\label{eq:QHJX0}%
\end{align}
where we have used the chain rule.

Let us find the total differential of $p_{0}$ remembering we are regarding it
as a parameter independent of time so that
\begin{align}
\frac{dp_{0}}{dt}=\frac{\partial^{2}S}{\partial x_{0}\partial t}%
+\frac{\partial^{2}S}{\partial x\partial x_{0}}\dot x=0.\label{eq:B}%
\end{align}
Subtracting (\ref{eq:B}) from (\ref{eq:QHJX0}), we get
\begin{align*}
\frac{\partial^{2}S}{\partial x\partial x_{0}}\left( \frac{\partial H^{\psi}%
}{\partial p}-\dot x\right) =0.
\end{align*}
Since we have assumed that $\frac{\partial^{2} S}{\partial x\partial x_{0}}%
\ne0$ the first of Hamilton's equations emerges
\begin{align}
\dot x=\frac{\partial H^{\psi}}{\partial p}.\label{eq:Hx}%
\end{align}


To obtain the second equation, we differentiate equation (\ref{eq:QHJ}) with
respect to $x$ and find
\begin{align}
\frac{\partial^{2}S}{\partial x\partial t}+\frac{\partial H^{\psi}}{\partial
x}+\frac{\partial H^{\psi}}{\partial p}\frac{\partial^{2}S}{\partial x^{2}%
}=0.\label{eq:C}%
\end{align}
Introducing the canonical momentum $p(t)=\nabla_{x}S(x(t),x_{0};t,t_{0})$ and
differentiating with respect to $t$ we obtain
\begin{align*}
\frac{\partial^{2}S}{\partial x\partial t}=\dot p-\frac{\partial^{2}%
S}{\partial x^{2}}\dot x.
\end{align*}
Thus equation (\ref{eq:C}) can be written in the form
\begin{align*}
\dot p(t)-\frac{\partial^{2}S}{\partial x^{2}}\dot x+\frac{\partial H^{\psi}%
}{\partial x}+\frac{\partial H^{\psi}}{\partial p}\frac{\partial^{2}%
S}{\partial x^{2}}=0.
\end{align*}
Taking into account Hamilton's first, we find Hamilton's second equation
\begin{align}
\dot p(t)=-\frac{\partial H^{\psi}}{\partial x}.\label{eq:Hp}%
\end{align}
Hamilton's equations (\ref{eq:Hx}) and (\ref{eq:Hp}) will then give us an
ensemble of trajectories from the equations (\ref{eq:DiracPs}) that Dirac
assumed could be used to construct quantum trajectories.

The question therefore remains, ``What is the form of $H^{\psi}$?" The answer
has been provided by Bohm~\cite{db52}. What he actually showed in his original
paper was that if we consider the real part of the Schr\"{o}dinger equation
under polar decomposition of the wave function $\psi= R\exp[i S]$, we find the
equation
\begin{align}
\frac{\partial S(x,t)}{\partial t}+\frac{(\nabla S(x,t))^{2}}{2m} +Q^{\psi
}(x,t) +V(x)=0.\label{eq:A3}%
\end{align}
This equation is identical in form to the classical Hamilton-Jacobi equation
except that it contains an additional term, namely the quantum potential
energy $Q^{\psi}(x,t) $. In other words this suggests that we identify
\begin{align*}
H^{\psi}=H + Q^{\psi}%
\end{align*}
where $Q^{\psi}$ is given by
\begin{align}
Q^{\psi}(x,t)=-\frac{1}{2m}\frac{\nabla^{2}R(x,t)}{R(x,t)}.\label{eq:QP}%
\end{align}
A more detailed discussion of this whole approach will be found in de
Gosson~\cite{mdg01}.

Before going on to discuss in more detail the mathematical background to this
 approach and its relation to Dirac's proposals, we must make a point of
clarification. Notice that the function $S_{\epsilon}(x,x^{\prime})$
introduced in equation (\ref{eq:DBprop}) is a two point function, namely a
propagator, while the Bohm approach emerges from a one-point function, namely
the wave function. This may not be a problem since the propagator
$K(x,x^{\prime};t,t^{\prime})=\psi(x,t)$ is the wave function, being simply
the probability amplitude to get to $(x,t)$ no matter what the initial point
is~\cite{rfah65}. Let us explore this relation in more detail.

\subsection{Weak Values and Bohm Trajectories}

Although the quantum Hamilton-Jacobi equation has been used to calculate
trajectories~\cite{dbbh93, ph94}, their meaning has been controversial, and at
times they have even been regarded as meaningless~\cite{dz99}.  This is in spite
of the fact that as the quantum potential becomes negligible the quantum
trajectory deforms smoothly into a classical trajectory. There are two main
factors contributing to the rejection of the notion of a quantum trajectory. Firstly there is
the question of how we reconcile an uncertainty principle that arises from a
fundamentally non-commutative structure. The need for such a revolutionary
structure was made quite evident in the original work of Born, Dirac,
Heisenberg and Jordan~\cite{vdw67}, yet equation
(\ref{eq:A3}) seems to imply that we needn't concern ourselves with such
complications. Unfortunately this is an illusion and although the approach
does provide a useful, but partial insight into quantum phenomena, it is
important to realise that we need to understand how this view is compatible
with the underlying non-commutative structure. Secondly it has not previously
been possible to investigate and construct these trajectories experimentally.
With the appearance of weak values, this situation has now changed with the
realisation that

\begin{itemize}
\item The weak value is not, in general, an eigenvalue of the operator under consideration.

\item Weak values are complex numbers.

\item The real part of the weak value of the momentum operator is identical to
the momentum given in equation (\ref{eq:p1}) where $S_{\epsilon}$ is
identified with the phase of the wave function (the probability amplitude of
getting to a point $(x,t)$).

\item It is possible to measure weak values even though they are not
eigenvalues, opening up the possibility of experimentally investigating the
precise meaning of these trajectories.
\end{itemize}

Recall that the weak value of the momentum can be written as
\begin{equation}
\frac{\langle x|\hat P|\psi(t)\rangle}{\langle x|\psi(t)\rangle}=\nabla
_{x}S(x,t) -i\nabla\rho(x,t)/2\rho(x,t)
\end{equation}
where we have chosen the polar decomposition of the wave function with
$\rho(x,t)=|\psi|^{2}$. Notice that the real part of this weak value can be
written as
\begin{align}
\Re\langle\hat P\rangle_{w_{x_{i},\psi}}=\Re\left[ \frac{\langle x|\hat
P|\psi\rangle}{\langle x|\psi\rangle}\right] =\frac{\partial S(x)}{\partial
x}\label{eq:WVP}%
\end{align}
which suggests that there may be some connection with the $p_{i+1}$ appearing in
equation (\ref{eq:p1}). Notice also that the Dirac expressions emerge from a
two-point propagator $S_{\epsilon}(x,x_{0},t,t_{0})$, not from the phase of a
wave function. And what about equation (\ref{eq:p2}) and the imaginary part of
the weak value? Let us now look at these relations from another angle.

\subsection{Relation of Weak Values to Non-Commutativity}

Let us rewrite the expressions (\ref{eq:p1}) and (\ref{eq:p2}) in a different
way to open up a new investigation
\begin{equation}
p_{q}=\frac{\langle q|\overrightarrow P|Q\rangle}{\langle q|Q\rangle}%
\quad\mbox{and}\quad p_{Q}=\frac{\langle Q|\overleftarrow P|q\rangle}{\langle
Q|q\rangle}.
\end{equation}
If we now form the sum $p_{q} +p_{Q}$, we find
\begin{equation}
\left[ p_{q}+p_{Q}\right] =\left[ \frac{\langle q|\overrightarrow P|Q\rangle
}{\langle q|Q\rangle}+\frac{\langle Q|\overleftarrow P|q\rangle}{\langle
Q|q\rangle}\right]
\end{equation}
while the difference gives
\begin{equation}
[p_{q}-p_{Q}]=\left[ \frac{\langle q|\overrightarrow P|Q\rangle}{\langle
q|Q\rangle} -\frac{\langle Q|\overleftarrow P|q\rangle}{\langle Q|q\rangle
}\right] .
\end{equation}
If we change the notation $\ |q\rangle\rightarrow| x\rangle$ and
$|Q\rangle\rightarrow|\psi\rangle$ we find
\begin{align}
\frac{1}{2} \left[ \frac{\langle x|\overrightarrow P|\psi\rangle}{\langle
x|\psi\rangle}+\frac{\langle\psi|\overleftarrow P|x\rangle}{\langle
\psi|x\rangle}\right] =\frac{\partial S(x,t)}{\partial x}=p_{B}%
(x,t)\label{eq:PB1}%
\end{align}
while
\begin{align}
\frac{1}{2} \left[ \frac{\langle x|\overrightarrow P|\psi\rangle}{\langle
x|\psi\rangle}-\frac{\langle\psi|\overleftarrow P|x\rangle}{\langle
\psi|x\rangle}\right] =-\frac{i}{2\rho(x,t)}\frac{\partial\rho(x,t)}{\partial
x}=-ip_{o}(x,t)\label{eq:PO1}%
\end{align}
where we have written $\psi(x,t)=\sqrt{\rho(x,t)}\exp[iS(x,t)]$. Notice that
both these momenta are real.

We may identify $p_{B}(x,t)$ with the Bohm or local momentum, while
$p_{o}(x,t)$ can be identified with what Nelson~\cite{en66}
calls the osmotic momentum. The origin of the term `osmotic' has its roots in
Nelson's attempts to derive the Schr\"{o}dinger equation by considering a
quantum particle undergoing a diffusive Brownian-type motion. Since a
continuous derivative is ruled out in a stochastic motion, we have to
distinguish between a forward derivative and a backward derivative.

In a non-commutative structure, we must distinguish between a left and a right
translation, so that both momenta, (\ref{eq:PB1}) and (\ref{eq:PO1}), arise by
combinations of the left and right translations of the momentum operator. This
implies that the real and imaginary parts of a weak value result from the fact
that we have, at the fundamental level, a non-commutative structure and by
forcing this into a complex structure we have hidden some aspects of the
deeper structure.

\subsection{Some preliminary comments on the experimental situation}

In a way we could claim that Dirac had essentially anticipated weak values, a
fact that has already been pointed out by Salek, Schubert and
Wiesner\cite{ssrs14}.
It should be noted that the weak value of the momentum is identical to the
local momentum\cite{mb13}, a notion that
has a long history going back to Landau\cite{ll41} and London\cite{fl45} in
the early discussions of the superfluid
properties of liquid helium. Because the local momentum could not be
represented by a linear operator, London concluded that it was not a
legitimate quantum observable as its value could not be measured in the
standard way.

However that all changed when Wiseman\cite{hw03} argued that the local
momentum,
being a weak value, could be measured in a process that Aharonov, Albert and
Vaidman\cite{yada88} called a \textquotedblleft weak measurement".
The ideas lying behind the weak measurement were considerably clarified by
Duck, Stevenson and Sudarshan~\cite{idpsgs89}. Not only was the principle of a
weak value and its measurement found to be correct, but an actual experiment carried out
by
Kocsis \emph{et al.}\cite{skbbas11} demonstrated how the local momentum could
be
measured in the interference region of a two-slit set up using a very weak
electromagnetic source produced by a quantum dot. By measuring the weak value
of the transverse local momentum at various positions in the region of
interference, they were able to construct momentum flow lines, which resembled
the Bohm trajectories calculated by Philippidis \emph{et al.}\cite{cdcpbh79}
and therefore the flow lines were interpreted as \textquotedblleft photon
trajectories"\cite{skbbas11}.

Unfortunately this identification is not as straightforward as it seems at
first sight. The trajectories constructed by Philippidis \emph{et al.} were
based on the Schr\"{o}dinger equation, whereas photons must be described by a
quantised Maxwell field. Again what appears to be a straightforward
generalisation of the notion of trajectories for atoms to those of photons is
not possible for reasons pointed out by Bohm, Hiley and
Kaloyerou\cite{dbbhpk87, rfbh14}.
Nevertheless the experimental determination of weak values has been
demonstrated and experiments are in progress to measure weak values using
atoms which, if successful, will open up a new debate in this
area~\cite{jmpepb16}.
Let us therefore return to a discussion of the deeper mathematical structure
lying behind these investigations.

\section{The Non-commutative Phase Space}

\subsection{Connection Between Commutative and Non-commutative Phase Space}

Even a glance at equation (\ref{eq:A3}) shows that when the quantum potential
energy $Q^{\psi}$ is negligible and $S$ identified with the classical action,
we recapture the classical Hamilton-Jacobi equation. In other words, we change
a non-commutative structure into a commutative structure. In terms of the
argument that it is the covering group that determines the behaviour of
quantum phenomena, the action is the second term in the expansion of
$\exp[iS]$ (\ref{eq:DBprop}). For a detailed discussion of the relationship
between the classical action and the phase of the wave function see de
Gosson~\cite{mdg01}.

Can we see, in a simple geometric way, how the space and its cover are related
in a manner that helps with the understanding of the problem we are facing
here without going into technical details? For the purposes of this paper
formality is secondary, as a formal discussion already exists
elsewhere~\cite{Landsman}.

To this end let us start by considering two points in a configuration space.
Here we will simply write the coordinates of a single point as $(x,t)$. Let us
introduce a characteristic operator $\rho=|\psi\rangle\langle\psi|$, which in
our configuration space we write as
\begin{align*}
\rho(x',x,t)=\psi^*(x^{\prime},t)\psi(x,t).
\end{align*}
In $p$-space we write
\begin{align*}
\phi(p,t)=(2\pi)^{-1}\int\psi(x,t)e^{-ipx}dx
\end{align*}
so that
\begin{align*}
\rho(x',x,t)=\int\int\phi^{*}(p',t)e^{ -ip^{\prime}x^{\prime}}%
\phi(p,t)e^{ixp}dpdp^{\prime}.
\end{align*}
Let us now change coordinates and use
\begin{align*}
X=(x^{\prime}+x)/2\quad\eta=x^{\prime}-x\quad P=(p^{\prime}+p)/2\quad
\pi=p^{\prime}-p.
\end{align*}
Then
\begin{align*}
\rho(X,\eta,t)=(2\pi)^{-1}\int\int\phi^{*}(P-\pi/2,t)\phi(P+\pi/2,t)e^{iX\pi
}e^{i\eta P}d\pi dP
\end{align*}
which we can write as
\begin{align*}
\rho(X,\eta,t)=(2\pi)^{-1}\int F(X,P,t)e^{i\eta P}dP.
\end{align*}
Taking the inverse Fourier transform of $\rho(X,\eta,t)$ will then provide us
with a characteristic function of a process now unfolding in a phase space,
where $(X,P)$ are the coordinates, not of a particle, but of a region in
configuration space characterised by a mean coordinate, $X$, and difference
coordinate, $\eta$, and a mean momentum $P$ and a difference $\pi$. These
parameters provide a measure, in some limited sense, of the size of the region
to which the energy of movement that is called \emph{a particle} is confined.

\subsection{Tangent Groupoids}

\label{sec:3.2}

It might seem that the introduction of a pair of points in configuration space
is arbitrary. However a deeper analysis of the underlying non-commutative
structure and its relation to the emergence of classical phase space has
helped to clarify the geometric structure underlying quantum phenomena.
Recently developed mathematical techniques  (Connes~\cite{ac90} and 
Landsman~\cite{Landsman}) based on asymptotic morphisms
between $C^{*}$-algebras show the deep relations between the Moyal algebra, an
algebra of a non-commutative phase space, and the Poisson algebra of classical
phase space. One of the key ingredients of this approach is the tangent
groupoid, a technique unfamiliar to the physics community so we will discuss
this approach in a subsequent paper. We introduce these ideas here merely to
indicate that there is a much richer structure underlying quantum phenomena
that is just beginning to be revealed with the exploration of weak values. For
preliminary details see Hiley~\cite{bh11a}.

\subsection{Return to Dirac}

We now consider the Dirac proposal of finding quantum trajectories. Notice
first that the two points, $(x,x^{\prime})$, chosen were conjugate points. The
corresponding operators of the mean variables $(X,P)$ then satisfy the
commutator $[X,P]=0$, i.e. this pair of operators are commutative and
therefore can have simultaneous eigenvalues which means a trajectory based on
those operators can be well defined.

To carry the comparison further we have to note that Dirac also includes a
pair of times $(t,t^{\prime})$, whereas we have one time.
In section \ref{sec:timedev} we will show how to generalise this approach to
consider pairs of space-time points. A more general and detailed discussion
will be found in Hiley~\cite{bh11, bh15b}.

Replacing the notion of a particle by a region of active energy may, at first
sight seem quite bizarre, but remember we are faced with a non-commutative
phase space and this must of necessity include novel features. One of these is
that the ordinary inner product must be replaced by a more general,
non-commutative product that is translation and symplectic equivariant,
associative, and non-local. There already exists a product with these
properties, namely, the well known Moyal star product~\cite{jgbjv88} to which
we have already referred. A more detailed discussion of the relationship
between the Moyal structure
and the algebraic approach has been discussed in Hiley~\cite{bh15}.

A further consequence of this relationship follows by performing a Fourier
transformation on the characteristic operator to show it can be written in the
form
\begin{align*}
F(X,P,t)=(2\pi)^{-1}\int\psi^{*}(X-\eta/2,t)e^{-i\eta P}\psi(X+\eta/2,t)d\eta.
\end{align*}
This will be immediately recognised as the Wigner function, a density matrix
introduced for a different problem than the one we are discussing
here~\cite{ew32}. For us it is the propagator of the time evolution of the
process. There is no necessity to regard this function as a `probability
distribution' as is done in quasi-classical quantum mechanics. We regard this
as providing a weighting function for each operator under consideration and
therefore no problem arises when it takes on negative values.

\subsection{Connection with the Orthogonal Clifford Algebra}

In section \ref{sec:chj} we pointed out that quantum phenomena could be
accounted for by going to the covering group of the symplectic group. This
brings out the close geometric relation between the classical and the quantum
behaviour. As we have already remarked a similar situation arises in the more
familiar case of spin. Here the spin group, $SU(2)$, is the covering group of
the rotation group, $SO(3)$. To analyse this structure, we have to go to the
Clifford algebra, which, in this case, is a non-commutative algebra. All
physicists are familiar with the anti-commutative structure of the Pauli
$\sigma$ matrices and the Dirac $\gamma$ matrices but their use as geometric
entities is novel.

These matrices are merely the representations of the generators of the
respective Clifford algebras. The advantage of using the Clifford algebra is
that the properties of the covering group can be obtained from the algebra
itself. Indeed the covering group is the Clifford group which appears as a
group of inner-automorphisms of the algebra and it turns out that one can work
completely from within the algebra, with no need to represent properties in an
abstract Hilbert space so that the wave function can be dispensed with. The wave function is
not essential and has merely been introduced as an algorithm for calculating
the probable outcome of a given system.

\subsection{Non-commutative Time Development Equations}

\label{sec:timedev}

In the context of a non-commutative algebra, it is important, once again, to
remember that we must distinguish between left and right translations. If we
use the ket $|\psi\rangle$ to specify the state of the system then only left
translations are possible. Furthermore it does not capture the fact that the
wave function is a special case of a propagator as Feynman suggests. Therefore
it was proposed that to obtain a description that allows both left and right
time translations on an equal basis, we need to make a generalisation of the
density operator, $\rho_{\psi,\phi}= |\psi\rangle\langle\phi|$. This operator
characterises the process under investigation, and can be used in the special
case of $\rho_{\psi,\psi}$, characterising the so-called `state of the
system'. This also has the advantage of allowing a straightforward
generalisation to mixed states. We will only be concerned with pure states in
this paper when $\rho^{2}=\rho$.

We will now assume that the equation for the left time translation is
\begin{align*}
i\frac{\partial|\psi\rangle}{\partial t}=\overrightarrow H|\psi\rangle,
\end{align*}
while the right time translation is governed by the equation
\begin{align*}
-i\frac{\partial\langle\psi|}{\partial t}=\langle\psi|\overleftarrow H.
\end{align*}
We have seemed not to have gained anything new compared with the standard approach
because, surely this is simply writing down the Schr\"{o}dinger equation and
its complex conjugate equation and therefore apparently adds no new
information. However when we consider the Pauli and Dirac equations, the left
and right translations do not have such a simple relationship~\cite{bhbc12}.

To see what new information these two equations contain, let us first form
\begin{align*}
i\frac{\partial|\psi\rangle}{\partial t}\langle\psi|=(\overrightarrow
H|\psi\rangle)\langle\psi|\quad\mbox{and}\quad-i|\psi\rangle\frac
{\partial\langle\psi|}{\partial t}=|\psi\rangle(\langle\psi|\overleftarrow H).
\end{align*}
If we now add and subtract these two equations we obtain the following two
equations, the first being
\begin{align}
i\frac{\partial\rho}{\partial t}=[H,\rho]_{-}\label{eq:hl}%
\end{align}
which is recognised as Heisenberg's equation for the time development of the
density operator. In the classical limit it becomes the Liouville equation
describing the conservation of probability. The second equation, arising from
the difference between the two equations, gives
\begin{align}
i\left( |\psi\rangle\frac{\overleftrightarrow\partial}{\partial t}\langle
\psi|\right) =[H,\rho]_{+}\label{eq:engcon}%
\end{align}
where
\begin{align*}
\left( |\psi\rangle\frac{\overleftrightarrow\partial}{\partial t}\langle
\psi|\right) =\left( \frac{\partial|\psi\rangle}{\partial t}\right)
\langle\psi|- |\psi\rangle\left( \frac{\partial\langle\psi|}{\partial
t}\right) .
\end{align*}

It should be remarked in passing that these equations are quite general and
have been used in the case of the Pauli and Dirac equations~\cite{bhbc12}. It
should also be noticed that in the two equations, (\ref{eq:hl}) and
(\ref{eq:engcon}), the quantum potential does not appear. For the full
generalisation the kets and bras must be replaced by appropriate elements of
the minimal left and right ideals in their respective algebras but we will not
discuss this approach further here. The details can be found in Hiley and
Callaghan~\cite{bhbc12} where it is
shown how these elements can be represented by matrices.

To link up with the Schr\"{o}dinger equation in its usual form, we must treat
$|\psi\rangle$ as an element in the algebra, which can be polar decomposed,
$\hat\Psi=\hat R\exp[i\hat S]$, and then inserted into equation
(\ref{eq:engcon}) to find
\begin{align*}
2\rho\frac{\partial\hat S}{\partial t}+[H,\rho]_{+}=0.
\end{align*}
This is just an equation for the conservation of energy that was first
introduced by Dahl~\cite{jd83}.
However if these equations are projected into a representation $|a\rangle$, we
find the equations
\begin{align*}
i\frac{\partial P(a)}{\partial t}+\langle[\rho,H]_{-}\rangle_{a}=0
\end{align*}
and
\begin{align*}
2P(a)\frac{\partial S}{\partial t}+\langle[\rho,H]_{+}\rangle_{a}=0.
\end{align*}
If we choose the $x$-representation, we find
\begin{align*}
\frac{\partial P}{\partial t}+\nabla.\left( P\frac{\nabla S_{x}}{m}\right) =0.
\end{align*}
Here $P(x,t)$ is the probability of finding the particle at $(x, t)$ and
$S_{x} $ is the phase of the wave function in the $x$-representation. The
second equation becomes
\begin{align*}
\frac{\partial S_{x}}{\partial t}+\frac{1}{2m}\left( \frac{\partial S_{x}%
}{\partial x}\right) ^{2}+\frac{Kx^{2}}{2}-\frac{1}{2mR_{x}}\left(
\frac{\partial^{2}R_{x}}{\partial x^{2}}\right) =0.
\end{align*}
Here the quantum potential appears for the first time. Thus the quantum
potential emerges only when the time development equation is projected into a
specific representation, in this case, the $x$-representation. Notice also
that, on polar decomposition of the wave function, the two equations,
(\ref{eq:hl}) and (\ref{eq:engcon}), produce separately the real and the
imaginary parts of the Schr\"{o}dinger equation as two real but coupled equations.

If we were to choose to project these equations into the $p$-representation,
we would obtain a different quantum potential. In fact the energy conservation
equation now becomes
\begin{align*}
\frac{\partial S_{p}}{\partial t}+\frac{p^{2}}{2m}+\frac{K}{2}x^{2}_{r}%
-\frac{K}{2R_{p}}\left( \frac{\partial^{2}R_{p}}{\partial p^{2}}\right) =0.
\end{align*}
Notice here we use $x_{r}=-\nabla_{p}S_{p}$, rather than $p_{x}=\nabla
_{x}S_{x}$. A more detailed discussion of the consequences of a quantum
potential appearing in the $p$-representation can be found in Brown and
Hiley~\cite{mbbh00}.

In this context the appearance of two projections at first sight seems rather
strange and, for some, certainly unwelcome. However it restores the $x-p$
symmetry, the perceived lack of which Heisenberg~\cite{wh58} originally used
as a criticism
against the Bohm approach, but at the same time it destroys the comfortable
intuitive form of the Bohm approach as the quantum process unfolding in an
\emph{a priori} given space-time. This opens up more radical approaches of the
type that Bohm was already aware and was actively investigating~\cite{db80}.
In this paper we will not go into the interpretation of these results. Those
interested will find details in ~\cite{bh16}.

Before concluding there are several features of this approach that should be
noted. The two time-development equations (\ref{eq:hl}) and (\ref{eq:engcon})
do not contain the complex wave function but correspond, in fact, to the
imaginary and real parts respectively of the Schr\"{o}dinger equation.
Secondly by replacing the bras and kets by what Dirac~\cite{pd47} calls
standard bras and standard kets, it can be shown that all the elements are
contained within the algebra itself. An external Hilbert space is not needed.
It is important to note this because interpretations based solely on Hilbert
space vectors miss the deeper mathematical structure which is in need of a
radically new interpretation. Thirdly, this approach does not require
retro-causation which is very much in fashion at the time of writing.
Fourthly, the Bohm approach is deeply imbedded in the quantum formalism and
the search for potential disagreements with the results of experiments
predicted by the standard approach is futile.

\section{Acknowledgements}

Maurice de Gosson has been supported by a research grant from the Austrian
Research Agency FWF (Projektnummer P27773--N13). Basil J. Hiley would like to
thank the Fetzer Franklin Fund of the John E. Fetzer Memorial Trust for their
support.



\bibliographystyle{plain}
\bibliography{myfile}
{}

\end{document}